\documentclass[twocolumn,amsmath,showpacs,amssymb,prl,aps]{revtex4-1}
\usepackage{graphicx,color,SIunits,bm} 
\newcommand{\Beta}{{\rm B}}

\begin{document}

\title{Quantum effects in the thermoelectric power factor of
  low-dimensional semiconductors}
 
\author{Nguyen T. Hung$^{1}$}
\email{nguyen@flex.phys.tohoku.ac.jp}

\author{Eddwi H. Hasdeo$^{1}$}
\author{Ahmad R. T. Nugraha$^{1}$}
\author{Mildred S. Dresselhaus$^{2,3}$}
\author{Riichiro Saito$^{1}$}

\affiliation{$^1$Department of Physics, Tohoku University, Sendai
  980-8578, Japan \\ $^2$Department of Electrical Engineering,
  Massachusetts Institute of Technology, Cambridge, MA 02139-4307,
  USA\\ $^3$Department of Physics, Massachusetts Institute of
  Technology, Cambridge, MA 02139-4307, USA}

\begin{abstract}
  We theoretically investigate the interplay between the confinement
  length $L$ and the thermal de Broglie wavelength $\Lambda$ to
  optimize the thermoelectric power factor of semiconducting
  materials.  An analytical formula for the power factor is derived
  based on the one-band model assuming nondegenerate semiconductors to
  describe quantum effects on the power factor of the low dimensional
  semiconductors. The power factor is enhanced for one- and
  two-dimensional semiconductors when $L$ is smaller than $\Lambda$ of
  the semiconductors. In this case, the low-dimensional semiconductors
  having $L$ smaller than their $\Lambda$ will give a better
  thermoelectric performance compared to their bulk counterpart. On
  the other hand, when $L$ is larger than $\Lambda$, bulk
  semiconductors may give a higher power factor compared to the lower
  dimensional ones.
\end{abstract}

\date{\today}
\maketitle

Thermoelectricity is a promising technology to improve the renewable
energy performance through conversion of waste heat into electric
energy~\cite{heremans13-thermo,vining09-thermo}. The efficiency of a
solid-state thermoelectric power generator is usually evaluated by the
dimensionless figure of merit, $ZT = S^2 \sigma\kappa^{-1} T$, where
$S$ is the Seebeck coefficient, $\sigma$ is the electrical
conductivity, $\kappa$ is the thermal conductivity, and $T$ is the
absolute temperature.  A fundamental aspect in the research of
thermoelectricity is the demand to maximize the $ZT$ value by having
large $S$, high $\sigma$, and low $\kappa$.  
However, since $S$, $\sigma$ and $\kappa$ are generally
interdependent, it has always been challenging for researchers to find
materials with $ZT > 2$ at room temperature~\cite{majumdar04-thermo}.
Huge efforts have been dedicated to reduce $\kappa$ using
semiconducting materials with low-dimensional structures, in which
$\kappa$ is dominated by phonon heat transport.  For example, recent
experiments using Si nanowires have observed that $\kappa$ can be
reduced below the theoretical limit of bulk Si ($0.99$ W/mK) because
the phonon mean free path is limited by boundary scattering in
nanostructures~\cite{boukai08-silicon,hochbaum08-silicon}.  In these
experiments, the reduction of the semiconducting nanowire diameter is
likely to achieve a large enhancement in thermoelectric efficiency
with $ZT > 1$ at room temperature~\cite{boukai08-silicon,
  hochbaum08-silicon}.  The success in reducing $\kappa$ thus leads to
the next challenge in increasing the thermoelectric power factor
$PF=S^2 \sigma$.

The importance of maximizing the $PF$ can be recognized from the fact
that when the heat source is unlimited, the $ZT$ value is no longer
the only one parameter to evaluate the thermoelectric efficiency.  In
this case, the output power density $Q$ is also important to be
evaluated~\cite{liu15-power,liu16-power}.  The $PF$ term appears in
the definition of $Q$, particularly for its maximum value, $Q_{\rm
  max}=PF(T_h-T_c)^2/4h_{\ell}$, where $T_h$, $T_c$, and $h_{\ell}$
are the hot side temperature, cold side temperature, and the length
between the hot and the cold sides (called the leg length),
respectively.  Since the term $(T_h-T_c)^2/4h_{\ell}$ is given by the
boundary condition, $Q$ is mostly affected by $PF$.  Here we mention
the definition of $Q$ because some materials show high $ZT$ but low
thermoelectric performance due to their small $Q$.  For example, Liu
\emph{et al.}  has compared two materials: PbSe (with maximum values
of $ZT = 1.3$, $PF = 21$ \micro W/cmK$^{2}$) and
Hf$_{0.25}$Zr$_{0.75}$NiSn ($ZT = 1$, $PF = 52$ \micro W/cmK$^{2}$) at
$T_h=500$ \celsius\ and $T_c=50$ \celsius\ with a leg length
$h_{\ell}=2$ mm~\cite{liu16-power}.  Their calculation showed that
PbSe (Hf$_{0.25}$Zr$_{0.75}$NiSn) has thermoelectric efficiency of
about $11\%$ ($10\%$), while its output $Q$ is about
$5.4~\textrm{W/cm}^2$ ($14.4~\textrm{W/cm}^2$).  From this
information, we can see that although PbSe has a larger $ZT$, its
output power is smaller than Hf$_{0.25}$Zr$_{0.75}$NiSn.  Therefore,
increasing the $PF$ value is important to enhance not only $ZT$ but
also $Q$ for power generation applications.  We thus would like to
consider the issue of maximizing $PF$ as the main topic of the present
work.

Of several methods to increase the $PF$ value, the reduction of the
confinement length $L$, which is defined by the effective size of the
electron wave functions in the non-principal direction for
low-dimensional materials, such as the thickness in thin films and the
diameter in nanowires, might be the most straightforward technique,
since it was proven to substantially increase
$ZT$~\cite{hicks96-thermo,hochbaum08-silicon,poudel08-thermo,kim15-Bi}.
A groundbreaking theoretical study by Hicks and Dresselhaus in 1993
predicted that a decrease in $L$ can increase $PF$ and $ZT$ of
low-dimensional structures~\cite{hicks93-thermo-well,
  hicks93-thermo-wire}.  However, if we look at some previous works
more carefully regarding the subject of the effect of confinement on
the $PF$, there were some experiments which showed that the $PF$ of
one-dimensional (1D) Si nanowires is still similar to that of the 3D
bulk system~\cite{boukai08-silicon,hochbaum08-silicon}, while other
experiments on Bi nanowires show an enhanced $PF$ value compared to
its bulk state~\cite{kim15-Bi}.  These situations indicate that there
is another parameter that should be compared with $L$.  We will show
in this Letter that the thermal de Broglie wavelength $\Lambda$ is a
key parameter that defines quantum effects in thermoelectricity.  In
order to show these effects, we investigate the quantum confinement
effects on the $PF$ for typical low-dimensional semiconductors.  By
comparing $L$ with $\Lambda$, we discuss the quantum effects and the
classical limit on the $PF$, from which we can obtain an appropriate
condition to maximize the $PF$.

In this Letter, we give an analytical formula for the optimum $PF$
value which can show the interplay between the quantum confinement
length and the thermal de Broglie wavelength of semiconductors with
different dimensionalities.  We apply the one-band model with the
relaxation time approximation (RTA) to derive the analytical formula
for the $PF$ of nondegenerate semiconductors.  The justification for
the one-band model with the RTA was already given in some earlier
studies, which concluded that the model was accurate enough to predict
the thermoelectric properties of low dimensional semiconductors, such
as semiconducting carbon nanotubes (s-SWNTs)~\cite{nguyen15-thermo},
Bi$_2$Te$_3$ thin films~\cite{hicks93-thermo-well}, and Bi
nanowires~\cite{hicks93-thermo-wire,sun99-thermo}.  To obtain the $PF$
formula in this work, we use similar analytical expressions for the
Seebeck coefficient $S$ and the electrical conductivity $\sigma$ which
were derived in our previous paper~\cite{nguyen15-thermo}.  However,
compared with Ref.~\onlinecite{nguyen15-thermo}, there is a
modification to the definition of the relaxation time $\tau(E)$ that
we adopt in the present work, i.e., $\tau(E) = \tau_0(E/k_{\rm
  B}T)^r$, where $\tau_0$ is the relaxation time coefficient, $E$ is
the carrier energy, $k_B$ is the Boltzmann constant, $T$ is the
average absolute temperature, and $r$ is a characteristic exponent
determining the scattering mechanism.  In
Ref.~\onlinecite{nguyen15-thermo}, $\tau(E)$ was defined by $\tau(E) =
\tau_0 E^r$~\cite{lundstrom00-semiconductor,zhou11-thermo}, where we
considered only the case of $r = 0$ or constant relaxation time
approximation (CRTA) for discussing the Seebeck coefficients of
s-SWNTs.  Redefinition of $\tau (E) = \tau_0(E/k_{\rm B}T)^r$ is,
however, suitable for purposes of this work.

The Seebeck coefficient $S$ and the electrical conductivity $\sigma$
are given, respectively, by~\cite{nguyen15-thermo,nguyen16-tauD}
\begin{equation}
\label{eq:S1}
S=-\frac{k_{\rm B}}{q}\left(\eta-r-\dfrac{D}{2}-1\right),
\end{equation}
and
\begin{equation}
\label{eq:S2}
\sigma =\frac{4 q^2 \tau_0
  \left(r + \frac{D}{2}\right) (k_{\rm B}T)^{D/2} \Gamma(r+\frac{D}{2})}{D\
  L^{3-D}(2\pi)^{D/2}\hbar^D\Gamma(\frac{D}{2})}(m^*)^{D/2-1}e^\eta,
\end{equation}
where $D = 1,2,~{\rm or}~3$ denotes the dimension of the material (1D,
2D, or 3D systems), $q=\pm e$ is the unit carrier charge, $m^*$ is the
effective mass of electrons or holes, $L$ is the confinement length
for a particular material dimension, $\Gamma(p)=\int_0^\infty
x^{p-1}e^{-x} dx$ is the Gamma function, $\eta=\zeta/k_{\rm B}T$ is
the reduced chemical potential (while $\zeta$ is defined as the
chemical potential measured from the top of the valence energy band in
a p-type semiconductor), $k_{\rm B}$ is the Boltzmann constant, and
$\hbar$ is Planck's constant.  Note that for an n-type semiconductor,
we can redefine $\eta$ or $\zeta$ to be measured from the bottom of
the conduction band, while the formulas for $S$ and $\sigma$ remain
the same.  From Eqs.~\eqref{eq:S1} and~\eqref{eq:S2}, the
thermoelectric power factor can be written as
\begin{equation}
\label{eq:S3}
PF\equiv S^2 \sigma=A(\eta-C)^2 e^\eta,
\end{equation}
where $A$ (in units of $\textrm{W/mK}^2$) and $C$ (dimensionless) are
given by
\begin{equation}
\label{eq:S4}
A=\frac{4\tau_0 k_{\rm B}^2}{L^3 m^*}
\left(\frac{L}{\Lambda}\right)^D
\frac{\left(r+\frac{D}{2}\right)\Gamma
  \left(r+\frac{D}{2}\right)}{D\ \Gamma \left(\frac{D}{2}\right)},
\end{equation}
and $C=r+D/2+1$, respectively.  In Eq.~\eqref{eq:S4}, the thermal de
Broglie wavelength is defined by
\begin{equation}
\label{eq:S5}
\Lambda=(2\pi\hbar^2/k_{\rm B} T m^*)^{1/2}
\end{equation}
which is a measure of the thermodynamic
uncertainty for the localization of a particle of mass $m^*$ with the
average thermal momentum
$\hbar(2\pi/\Lambda)$~\cite{silvera97-wavelength}.

For a given $\tau(E)$, the carrier mobility is defined by
\begin{equation}
\label{eq:S6}
\mu=\frac{q\langle\langle\tau(E)\rangle\rangle}{m^*}, 
\end{equation}
where
\begin{equation}
\label{eq:S7}
\langle\langle\tau(E)\rangle\rangle\equiv
\frac{\langle E\tau(E)\rangle}{\langle E\rangle}=
\tau_0\frac{\Gamma\left(\frac{5}{2}+r\right)}{\Gamma \left(\frac{5}{2}\right)},
\end{equation}
and $\langle x \rangle = \int_0^\infty x e^{-E/k_{\rm B}T}dE$ in
Eq.~\eqref{eq:S7} is a canonical average of $x$.  From
Eqs.~\eqref{eq:S4},~\eqref{eq:S6} and~\eqref{eq:S7}, the term $A$ of
the power factor can be rewritten as
\begin{equation}
\label{eq:S8}
A=\frac{4\mu k_{\rm B}^2}{qL^3 }
\left(\frac{L}{\Lambda}\right)^D
\frac{\left(r+\frac{D}{2}\right)
  \Beta\left(r,\frac{5}{2}\right)}{D\ \Beta\left(r,\frac{D}{2}\right)},
\end{equation}
where $\Beta(x,y)=\Gamma(x)\Gamma(y)/\Gamma(x+y)$ is the
Beta function.  We can now determine the optimum power factor as a
function of $\eta$ from Eq.~\eqref{eq:S3} by solving ${\rm d}(PF)/{\rm
  d}\eta=0$.  The optimum power factor, $PF_{\rm opt}$, is found to be
\begin{equation}
\label{eq:S9}
PF_{\rm opt}=\frac{16\mu k_{\rm B}^2}{qL^3}
\left(\frac{L}{\Lambda}\right)^D
\frac{\left(r+\frac{D}{2}\right)\Beta\left(r,\frac{5}{2}\right)}{D\ \Beta
  \left(r,\frac{D}{2}\right)}e^{r+D/2-1},
\end{equation}
whereas the corresponding value for the reduced (dimensionless)
chemical potential is $\eta_{\rm opt} =r+D/2-1$.

\begin{figure}[t!]
  \centering
  \includegraphics[clip,width=8cm]{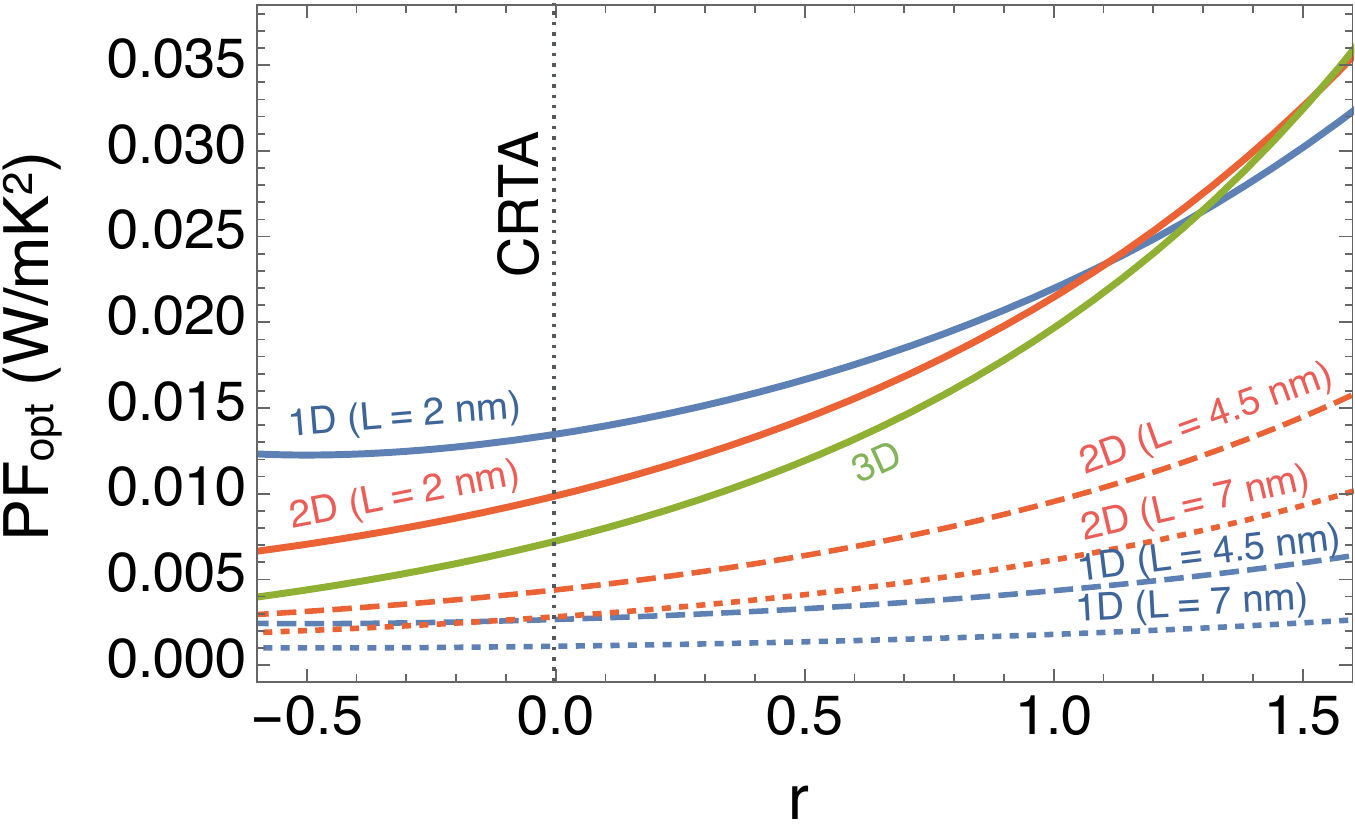}
  \caption{\label{fig:scatering} (Color online) Optimum power factor
    $PF_{\rm opt}$ as a function of characteristic exponent $r$ for
    the 1D, 2D, and 3D systems.  The thermal de Broglie wavelength is
    set to be $\Lambda=4.5~\textrm{nm}$ (for n-type Si) and the
    mobility is $\mu=420~\textrm{cm}^2/\textrm{Vs}$.  The confinement
    length $L$ is varied for the 1D and 2D systems, each for $L =
    2~\textrm{nm}$, $L=\Lambda$ (4.5 nm), and $L = 7~\textrm{nm}$.
    The value of $r = 0$ corresponds to the constant relaxation time
    approximation (CRTA).}
\end{figure}

Next, we discuss some cases where $PF_{\rm opt}$ may be enhanced
significantly.  Figure~\ref{fig:scatering} shows $PF_{\rm opt}$ as a
function of the characteristic exponent $r$ for the 1D, 2D, and 3D
systems, in which the values of $r$ range from $-0.5$ to $1.5$ for
various scattering
processes~\cite{lundstrom00-semiconductor,zhou11-thermo}.  In these
examples, we consider a typical semiconductor, n-type Si, at room
temperature and high-doping concentrations on the order of $10^{18}$
cm$^{-3}$.  The thermal de Broglie wavelength and the carrier mobility
are set to be $\Lambda=4.5~\textrm{nm}$ and
$\mu=420~\textrm{cm}^2/\textrm{Vs}$, respectively.  We note that the
scattering time assumed under the CRTA corresponds to $r = 0$, and
thus $\langle\langle \tau(e)\rangle\rangle \equiv
\tau_0$~\cite{stradling70-scattering}.  As shown in
Fig.~\ref{fig:scatering}, $PF_{\rm opt}$ increases with increasing $r$
for all the 1D, 2D, and 3D systems. The effect of the characteristic
exponent $r$ on the 3D system is stronger than that of the 1D and 2D
systems.  Based on Eq.~\eqref{eq:S9} and Fig.~\ref{fig:scatering},
$PF_{\rm opt}$ increases with decreasing $L$ corresponding to the
confinement effect for the 1D and 2D systems. It is noted in
Fig.~\ref{fig:scatering} that $PF_{\rm opt}$ in the 3D system does not
depend on $L$ as shown in Eq.~\eqref{eq:S9} with $D=3$.  However, the
qualitative behaviour between $r$ and $PF_{\rm opt}$ is not much
affected by changing $L$ since $r$ and $L$ are independent of each
other in Eq.~\eqref{eq:S9}.

\begin{figure}[t!]
  \centering \includegraphics[clip,width=8.5cm]{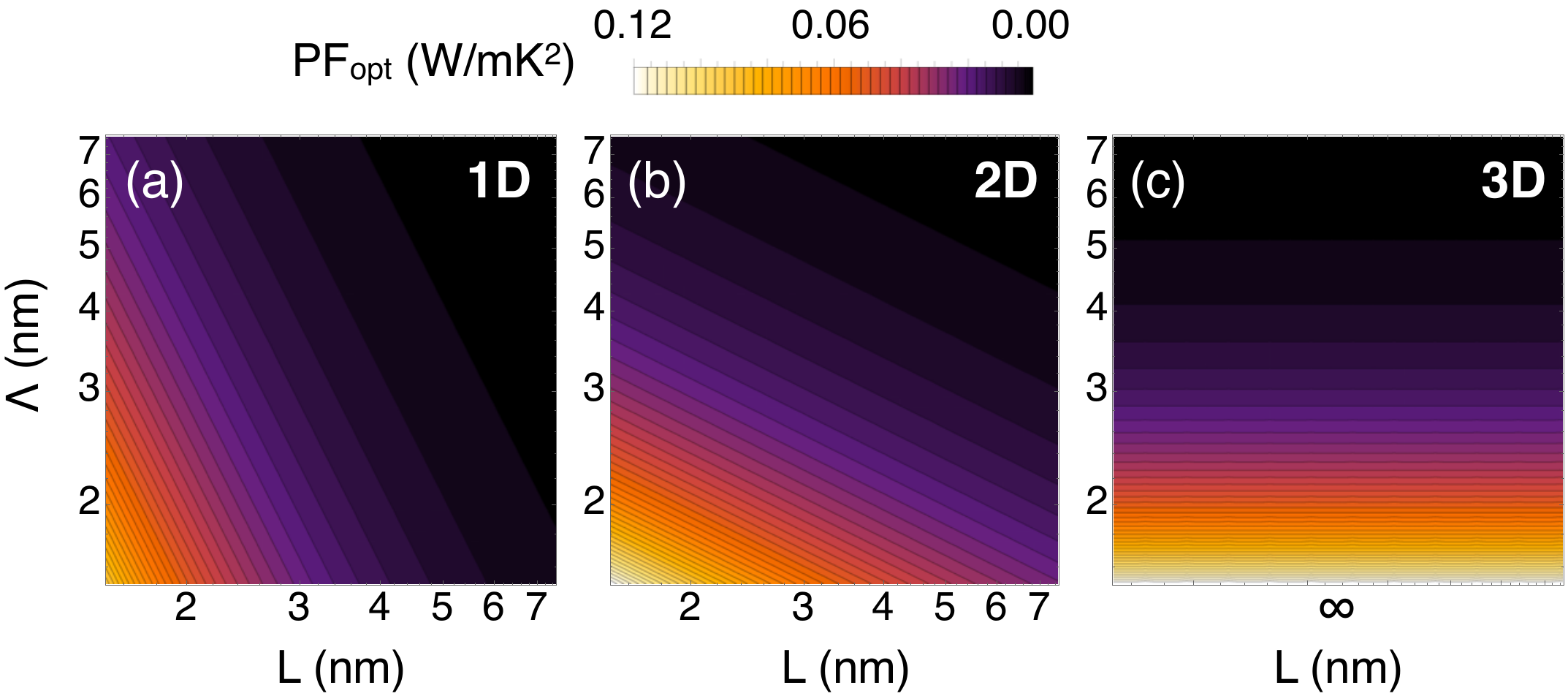}
  \caption{\label{fig:PF_L_G} (Color online) Optimum power factor
    $PF_{\rm opt}$ as a function of confinement length $L$ and thermal
    de Broglie wavelength $\Lambda$ plotted on a logarithmic scale for
    (a) 1D, (b) 2D, and (c) 3D systems.}
\end{figure}

Figure~\ref{fig:PF_L_G} shows $PF_{\rm opt}$ as a function of
confinement length $L$ and thermal de Broglie wavelength $\Lambda$ for
the 1D, 2D, and 3D systems.  The mobility is set to be $\mu =420$
cm$^2$/Vs for each system and the scattering rate may be proportional
to the density of final states (DOS).  By assuming proportionality of
the scattering rate with respect to the DOS, we obtain $r=+0.5$, $r=0$
and $r=-0.5$ for 1D, 2D, and 3D systems,
respectively~\cite{zhou11-thermo}.  Hereafter, we consider such
different $r$ values for the different dimensions.  The curves in
Figs.~\ref{fig:PF_L_G}(a) and (b) in particular show a $L^{-2}$ and
$L^{-1}$ dependence of $PF_{\rm opt}$ for 1D and 2D systems,
respectively [cf. Eq.~\eqref{eq:S9}].  These results are consistent
with the Hicks-Dresselhaus model~\cite{hicks93-thermo-well,
  hicks93-thermo-wire}.  In addition, in this Letter, we point out
that it is important to consider the dependence of $PF_{\rm opt}$ on
$\Lambda$.  For an ideal electron gas under a trapping potential, the
thermodynamic uncertainty principle may roughly be expressed as
$\Delta P/P \times \Delta V/V \geq (D^{3/2}/\sqrt{2\pi}){\Lambda}/L$,
where $P$ and $V$ are the pressure and volume of the system,
respectively~\cite{farag14-wavelength}.  The uncertainty principle
ensures that when the confinement length is comparable with the
thermal de Broglie wavelength, i.e., $L\leq
(D^{3/2}/\sqrt{2\pi}){\Lambda}$, the $P$ and $V$ cannot be treated as
commuting observables.  In this case, quantum effects play an
important role in increasing $PF_{\rm opt}$ for nanostructures.  For a
1D system [Fig.~\ref{fig:PF_L_G}(a)] $PF_{\rm opt}$ starts to increase
significantly when $L$ is much smaller than $\Lambda$, while for the
2D system [Fig.~\ref{fig:PF_L_G}(b)] $PF_{\rm opt}$ starts to increase
significantly when $L$ is comparable to $\Lambda$.  As for the 3D
system [Fig.~\ref{fig:PF_L_G}(c)], $PF_{\rm opt}$ increases with
decreasing $\Lambda$ for any $L$ values.  Therefore, a nanostructure
having both small $L$ and $\Lambda$ (while $L$ is also much smaller
than its $\Lambda$) will be the most optimized structure to enhance
$PF$.

\begin{figure}[t!]
  \centering \includegraphics[clip,width=8cm]{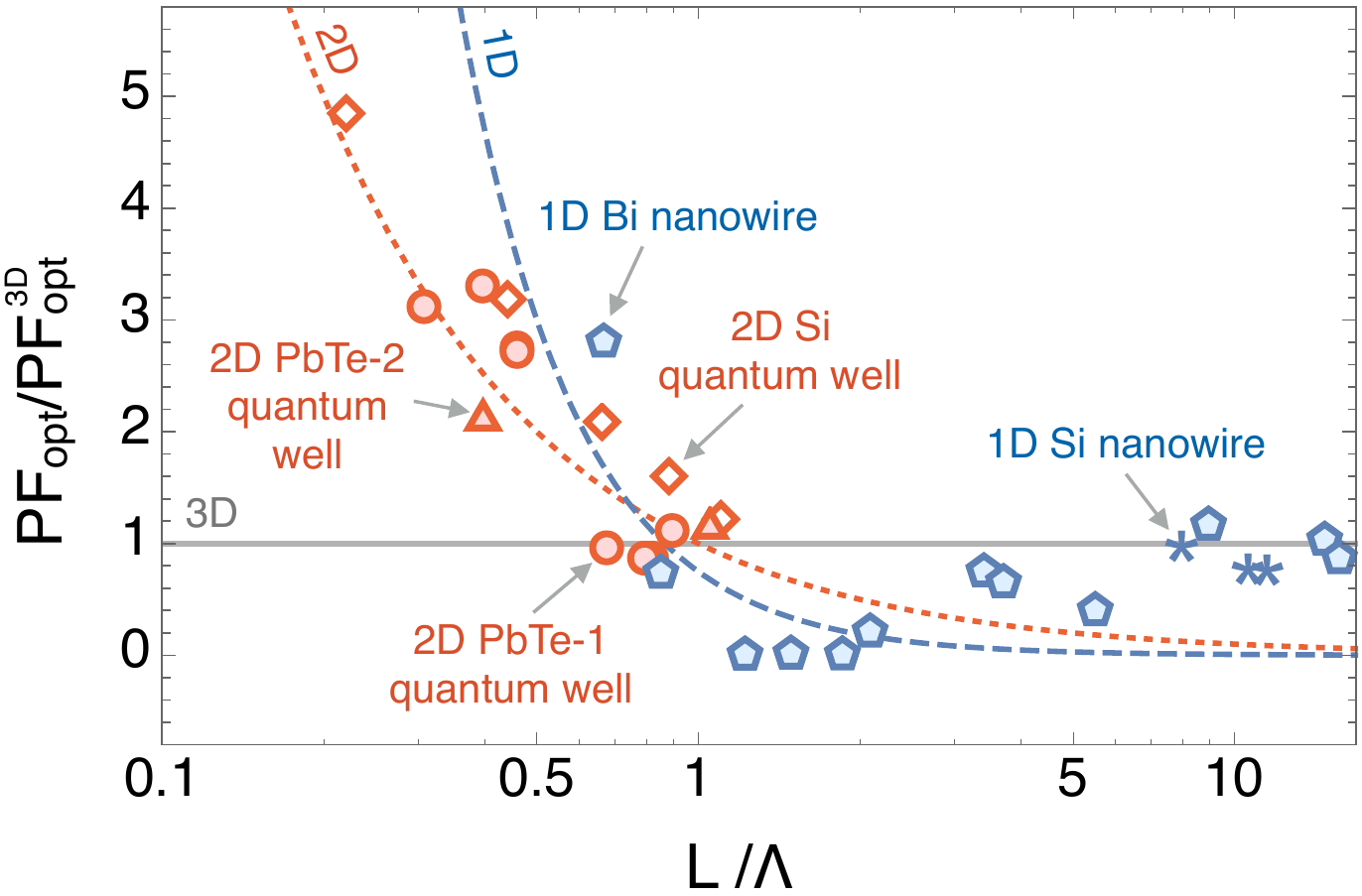}
  \caption{\label{fig:PF_LG} (Color online) $PF_{\rm opt}/PF_{\rm
      opt}^{\rm 3D}$ as a function of $L/\Lambda$ for different
    dimensions.  The $L/\Lambda$ axis is given using a logarithmic
    scale.  Theoretical results for 1D, 2D, and 3D systems are
    represented by dashed, dotted, and solid lines, respectively.
    Asterisks, pentagons, diamonds, circles, and triangles denote
    experimental results for 1D Si
    nanowires~\cite{hochbaum08-silicon}, 1D Bi
    nanowires~\cite{kim15-Bi}, 2D Si quantum wells~\cite{sun99-Si}, 2D
    PbTe--1 quantum wells~\cite{Harman96-PbTe}, and 2D PbTe--2 quantum
    wells~\cite{Hicks96-Si}, respectively.  For the experimental
    results, we set the thermal de Broglie wavelength of each material
    as: $\Lambda_{\rm Bi}= 32~{\rm nm}$, $\Lambda_{\rm Si}= 4.5~{\rm
      nm}$, and $\Lambda_{\rm PbTe}=5~{\rm nm}$.  We also have the
    following $PF$ values for 3D systems: $PF_{\rm Bi}^{\rm 3D}=
    0.002~{\rm W/mK^2}$~\cite{kim15-Bi}, $PF_{\rm Si}^{\rm 3D}=
    0.004~{\rm W/mK^2}$~\cite{weber91-Si}, $PF_{\rm PbTe-1}^{\rm 3D}=
    0.002~{\rm W/mK^2}$~\cite{Harman96-PbTe}, and $PF_{\rm
      PbTe-2}^{\rm 3D}= 0.003~{\rm W/mK^2}$~\cite{Hicks96-Si}.}
\end{figure}

Now we can compare our model with various experimental data.  In
Fig.~\ref{fig:PF_LG}, we show $PF_{\rm opt}$ as a function of
$L/\Lambda$ for different dimensions (1D, 2D, and 3D systems)
following Eq.~\eqref{eq:S9}.  The $PF_{\rm opt}$ values are scaled by
the optimum power factor of a 3D system, $PF_{\rm opt}^{\rm 3D}$.
From Eq.~\eqref{eq:S9}, we see that the ratio $PF_{\rm opt}/PF_{\rm
  opt}^{\rm 3D}$ merely depends on $L/\Lambda$ and $D$.  Hence, $PF$
from various materials can be compared directly with the theoretical
curves shown in Fig.~\ref{fig:PF_LG}.  The experimental data in
Fig.~\ref{fig:PF_LG} are obtained from the $PF$ values of 1D Bi
nanowires~\cite{kim15-Bi}, 1D Si nanowires~\cite{hochbaum08-silicon},
2D Si quantum wells~\cite{sun99-Si}, and two different experiments on
2D PbTe quantum wells labeled by PbTe--1 and
PbTe--2~\cite{Harman96-PbTe,Hicks96-Si}.  Here we use fixed parameters
for the thermal de Broglie wavelength of each material: $\Lambda_{\rm
  Bi}= 32~{\rm nm}$, $\Lambda_{\rm Si}= 4.5~{\rm nm}$, and
$\Lambda_{\rm PbTe}=5~{\rm nm}$.  We also set some $PF$ values for
bulk systems: $PF_{\rm Bi}^{\rm 3D}= 0.002~{\rm
  W/mK^2}$~\cite{kim15-Bi}, $PF_{\rm Si}^{\rm 3D}= 0.004~{\rm
  W/mK^2}$~\cite{weber91-Si}, $PF_{\rm PbTe-1}^{\rm 3D}= 0.002~{\rm
  W/mK^2}$~\cite{Harman96-PbTe}, and $PF_{\rm PbTe-2}^{\rm 3D}=
0.003~{\rm W/mK^2}$~\cite{Hicks96-Si}, which are necessary to put all
the experimental results into Fig.~\ref{fig:PF_LG}.

We find that the curves in Fig.~\ref{fig:PF_LG} demonstrate a strong
enhancement of $PF_{\rm opt}$ in 1D and 2D systems when the ratio $L /
\Lambda$ is smaller than one ($L < \Lambda$).  In contrast, if $L$ is
larger than $\Lambda$, the bulk 3D semiconductors may give a larger
$PF_{\rm opt}$ value than the lower dimensional semiconductors, as
shown in Fig.~\ref{fig:PF_LG} up to a limit of $L/\Lambda\approx2$.
We argue that such a condition is the main reason why an enhanced $PF$
is not always observed in some materials although experimentalists
have reduced the material dimensionality.  For example, in the case of
1D Si nanowires, where we have $\Lambda_{\textrm{Si}} \sim
4.5~\textrm{nm}$, we can see that the experimental $PF$ values in
Fig.~\ref{fig:PF_LG} are almost the same as the $PF_{\rm opt}^{\rm
  3D}$.  The reason is that the diameters (supposed to represent $L$)
of the 1D Si nanowires, which were about $36$--$52~\textrm{nm}$ in the
previous experiments~\cite{boukai08-silicon,hochbaum08-silicon}, are
still too large compared with $\Lambda_{\textrm{Si}}$.  It might be
difficult for experimentalists to obtain a condition of $L < \Lambda$
for the 1D Si nanowires.  In the case of materials having larger
$\Lambda$, e.g., Bi with $\Lambda_{\textrm{Bi}} \sim 32~\textrm{nm}$,
the $PF$ values of the 1D Bi nanowires can be enhanced at $L <
\Lambda$, which is already possible to achieve
experimentally~\cite{kim15-Bi}.  Furthermore, when $L\gg\Lambda$, it
is natural to expect that $PF_{\rm opt}$ of 1D and 2D semiconductors
resemble $PF_{\rm opt}^{\rm 3D}$ as shown by some experimental data in
Fig.~\ref{fig:PF_LG}.  It should be noted that, within the one-band
model, we do not obtain a smooth transition of $PF_{\rm opt}$ in
Fig.~\ref{fig:PF_LG} from the lower dimensional to the 3D
characteristics for large $L$ because we neglect contributions coming
from many other subbands responsible for the appearance of the 3D
density of states~\cite{cornett2011-powerfactor}.

So far, we have used the confinement length $L$ as an independent
parameter in Eq.~\eqref{eq:S9}.  It is actually possible to engineer
the confinement length in the same material.  For extremely thin films
or nanowires, $L$ is expressed by two components as $L=L_0+\Delta L$,
where $L_0$ is the thickness of the material and $\Delta L$ is the
size of the evanescent electron wavefunction beyond the surface
boundary.  Within the box of $L_0$ the electron wavefunction is
delocalized, approximated by the linear combination of plane waves,
while within $\Delta L$ the electron wavefunction is approximated by
evanescent waves.  For a single-layered material, e.g., a hexagonal
boron nitride (h-BN) sheet, $L_0 \approx 0$ so that $L \approx \Delta
L=0.333$ nm~\cite{lee10-thickness}.  As for ultra-thick 1D nanowires
or 2D thin films, we have $L \gg \Delta L$, and thus the confinement
length is mostly determined by the size of the material such as $L
\approx L_0$.  Creating a 1D channel from a 2D material by applying
negative gate voltages on two sides of the 2D material can be an
example to engineer the confinement length~\cite{hirayama89}.

We already see that the thermal de Broglie wavelength $\Lambda$
depends on the temperature and the effective mass for the material.
As given in Eq.~\eqref{eq:S5}, $\Lambda$ decreases ($\propto T^{-1/2}$
or ${m^*}^{-1/2}$) with increasing temperature $T$ or with increasing
effective mass $m^*$, which indicates that the $PF_{\rm opt}$
[$\propto (L/\Lambda)^D$ in Eq.~\eqref{eq:S9}] of nondegenerate
semiconductors would be enhanced at higher $T$ or at larger $m^*$
(smaller $\Lambda$).  This result is consistent with the experimental
observations for the $PF$ values of Si and PbTe, which are
monotonically increasing as a function of
temperature~\cite{hochbaum08-silicon,weber91-Si,sootsman08}.  It
should be noted that $\Lambda$ is not necessarily independent of $L$
and $D$ because the term $m^*$ may be altered by varying $L$ or by
changing $D$.  For example, based on the fitting in
Ref.~\onlinecite{sajjad09}, the effective masses of 1D Si nanowires
for $L$ within the interval of $2$--$40$~nm could change from $1.1~m_0$
to $0.8~m_0$, where $m_0$ is the free electron mass.  Meanwhile,
Ref.~\onlinecite{green90-jap} reported that bulk 3D Si has an
effective mass of about $1.09m_0$ at room temperature.  As a result,
we estimate that the change of $\Lambda$ is roughly about $5$--$10\%$ in
this case.  This fact might contribute to the small discrepancy
between the $PF$ values from our theory and those from experiments
since we set $\Lambda$ as a fixed quantity upon variation of $L$ in 1D
and 2D systems (see Fig.~\ref{fig:PF_LG}).  For the 3D system, the
theoretical values ($PF_{\rm Bi}^{\rm 3D}=0.0019~{\rm W/mK^2}$ and
$PF_{\rm Si}^{\rm 3D}=0.0044~{\rm W/mK^2}$) are in good agreement with
the experimental data ($PF_{\rm Bi}^{\rm 3D}=0.002~{\rm
  W/mK^2}$~\cite{kim15-Bi} and $PF_{\rm Si}^{\rm 3D}=0.004~{\rm
  W/mK^2}$~\cite{weber91-Si}).

In conclusion, we have shown that the largest power factor $PF$ values
might be obtained for low-dimensional systems by decreasing both the
confinement length $L$ and the thermal de Broglie wavelength $\Lambda$
while keeping $L < \Lambda$.  Depending on the materials dimension,
there is a different interplay between $L$ and $\Lambda$ to enhance
the power factor.  A simple analytical formula [Eq.~\eqref{eq:S9}]
based on the one-band model has been derived to describe the quantum
effects on the $PF$ in 1D, 2D, and 3D systems.  We would suggest to
experimentalists to be careful to check the trade-off between $L$ and
$\Lambda$ in order to enhance $PF$ for different dimensions of their
semiconductors.

N.T.H. and A.R.T.N acknowledge the Interdepartmental Doctoral Degree
Program for Multidimensional Materials Science Leaders in Tohoku
University.  R.S. acknowledges MEXT (Japan) Grants No. 25107005 and
No. 25286005. M.S.D acknowledges support from NSF (USA) Grant
No. DMR-1507806.

%\bibliographystyle{apsrev4-1}
%\bibliography{nguyen16-powerdim}

\begin{thebibliography}{30}
\bibitem [{ngu()}]{nguyen16-tauD}%
  \BibitemOpen
  \href@noop {} {}\bibinfo {note} {There is a difference in the exponent of the
  $k_B T$ terms in Eq. (2) of this work and Eq. (B3) in Ref. 13, i.e., $D/2$
  for this work and $D/2 + r$ for Ref. 13. This is a consequence of the
  definition of $\tau (E) = \tau_0 (E/k_B T)^r$ adopted in this work and
  $\tau(E) = \tau_0 E^r$ in Ref. 13.}\BibitemShut {Stop}%
\end{thebibliography}

%merlin.mbs apsrev4-1.bst 2010-07-25 4.21a (PWD, AO, DPC) hacked
%Control: key (0)
%Control: author (72) initials jnrlst
%Control: editor formatted (1) identically to author
%Control: production of article title (-1) disabled
%Control: page (0) single
%Control: year (1) truncated
%Control: production of eprint (0) enabled
%

\end{document}